\documentclass[useAMS,usenatbib,usegraphicx,a4paper,referee]{mn2e}
\newcommand {\aplt} {{\raise-.5ex\hbox{$\buildrel<\over\sim$}}} 
\newcommand{\priname}{TYC\,9486-927-1}
\newcommand{\secname}{2MASS\,J21265040$-$8140293}
\newcommand{\secnameshort}{2MASS\,J2126$-$8140}
\usepackage{hyperref}
\usepackage{graphicx}
 \title[A possible wide star-planet system]{A nearby young M dwarf with a wide, possibly planetary-mass companion\thanks{Based on observations made with the ESO/MPG 2.2m telescope at the La Silla Observatory under programme ID 090.A-9010.}}
 \author[N.R.\ Deacon et al.]{N.R.\ Deacon\thanks{E-mail:n.deacon2@herts.ac.uk}$^{1}$, J.E. Schlieder$^{2,3}$, S.J. Murphy$^4$\\
$^1$Centre for Astrophysics Research, University of Hertfordshire, College Lane, Hatfield, AL10 9AB, UK\\ 
$^2$NASA Postdoctoral Program Fellow, NASA Ames Research Center, Moffett Field, CA, USA\\
$^3$Max Planck Institute for Astronomy, Konigstuhl 17, Heidelberg, 69117, Germany\\
$^4$Research School of Astronomy and Astrophysics, Australian National University, Canberra, ACT 2611, Australia\\
}
 \begin{document}
 \date{}
 \pagerange{\pageref{firstpage}--\pageref{lastpage}} \pubyear{2013}
 \maketitle
 \label{firstpage}
 \begin{abstract} We present the identification of two previously known young objects in the solar neighbourhood as a likely very wide binary. \priname,\enspace an active, rapidly rotating early-M dwarf, and \secname,\enspace a low-gravity L3 dwarf previously identified as candidate members of the $\sim$45 Myr old Tucana Horologium association (TucHor). An updated proper motion measurement of the L3 secondary, and a detailed analysis of the pair's kinematics in the context of known nearby, young stars, reveals that they share common proper motion and are likely bound.  New observations and analyses reveal the primary exhibits Li 6708~\AA~absorption consistent with M dwarfs younger than TucHor but older than the $\sim$10\,Myr TW Hydra association yielding an age range of 10-45\,Myr. A revised kinematic analysis suggests the space motions and positions of the pair are closer to, but not entirely in agreement with, the $\sim$24 Myr old $\beta$ Pictoris moving group. This revised 10-45\,Myr age range yields a mass range of 11.6--15 M$_J$ for the secondary. It is thus likely \secnameshort\enspace is the widest orbit planetary mass object known ($>$4500AU) and its estimated mass, age, spectral type, and $T_{eff}$ are similar to the well-studied planet $\beta$ Pictoris b. Because of their extreme separation and youth, this low-mass pair provide an interesting case study for very wide binary formation and evolution.
  \end{abstract}
\section{Introduction}
Very wide orbit ($>$1000\,AU) extrasolar planets represent a currently small but previously unexpected population of massive gas giant companions to stars. To date four such objects have been discovered by direct imaging by a variety of routes. WD~0806-661B (\citealt{Luhman2011}; 2500\,AU 6--9\,M$_{J}$; \citealt{Luhman2012}) was discovered with a targeted observation of a nearby white dwarf, GU\,Psc\,b (\citealt{Naud2014}; 2000\,AU, 9--12\,M$_{J}$) was found with a targeted observation of a young, nearby moving group member,  SR12\,C by observations of a T~Tauri binary in $\rho$ Ophiuchus (\citealt{Kuzuhara2011}; 1100\,AU, 6--20\,M$_J$), whilst Ross\,458\,C (\citealt{Goldman2010}; 1160\,AU, 5--14\,M$_{J}$) was discovered in widefield survey data and then identified as having a common proper motion with its host binary.

In this work we present the identification of two previously known young objects in the solar neighbourhood, TYC 9486-927-1 and 2MASS J21265040-8140283, as a co-moving wide pair with a probable planetary mass secondary. During an examination of the literature we found that these two objects are separated by 217$"$ and have similar proper motions. Hence we attempted to better determine their properties to see if they were a likely young, bound system. Using revised astrometry and detailed kinematic analyses of nearby young stars and brown dwarfs, we have determined that this previously known young brown dwarf/free-floating planetary mass object and young low mass star are a likely widely separated bound pair. We also present new spectroscopic observations and re-examinations of literature and archival data to refine the age of the system and estimate that the secondary is likely planetary mass and may be the widest orbit exoplanet yet discovered.

TYC 9486-927-1 was observed by \cite{Torres2006} as part of the Search for Associations Containing Young stars (SACY) programme \citep{Torres2008}. They assigned a spectral type of M1 and measured a radial velocity of  $v_{rad}=8.7\pm4.6$\,km/s from 10 observations. The large uncertainty is likely due to the star's high rotational velocity ($v\sin i=43.5\pm1.2$\,km/s); suggesting it is either a single rapid rotator or a spectroscopic binary with blended lines. \priname\enspace also shows signs of activity in X-ray \citep{Thomas1998}, H$\alpha$ emission \citep{Torres2006} and the UV (using {\it GALEX} data from \citealt{Martin2005} we find $\log F_{FUV}/F_{J}$=-2.49, $\log F_{NUV}/F_{J}$=-2.11). \secnameshort\enspace is an L3 first identified by \cite{Reid2008} (although referencing \citealt{Cruz2009} as the discovery paper). Subsequently \cite{Faherty2013} classified it as a low gravity L3$\gamma$ (using the gravity classification system of \citealt{Cruz2009}). Recent VLT/ISAAC observations by \cite{Manjavacas2014} find it is a good match to the young, L3 companion CD-35 2722B \citep{Wahhaj2011}. These authors also used the spectral indices of \cite{Allers2013} to confirm that the \secnameshort\enspace is an L3 and shows low gravity spectral features. \cite{Manjavacas2014} also used the BT-Settl-2013 atmospheric models \citep{Allard2012} to derive $T_{eff}=1800\pm100$\,K, $\log g = 4.0\pm0.5$\,dex, albeit with better fits to super-solar metallicity models. \cite{Filippazzo2015} use photometry, a trigonometric parallax of 31.3$\pm$2.6\,mas (referenced to Faherty et al., in prep.) and evolutionary models to derive an effective temperature of 1663$\pm$35\,K. They also derived a mass of 23.80$\pm$15.19M$_J$ assuming a broad young age range of 10--150\,Myr. \cite{Gagne2014} listed \secnameshort\enspace as a high probability candidate member of Tucana-Horologium association (TucHor) but noted that its photometric distance would be in better agreement with its TucHor kinematic distance if it were an equal mass binary. 

\section{TYC 9486-927-1 and 2MASS~J21265040$-$8140293}
Our parameters for both components of this system are listed in Table~\ref{sys_sum}. Below we outline how these were derived.
\subsection{Astrometry of 2MASS~J21265040$-$8140293}
\begin{figure}
\begin{center}
\includegraphics[scale=0.5]{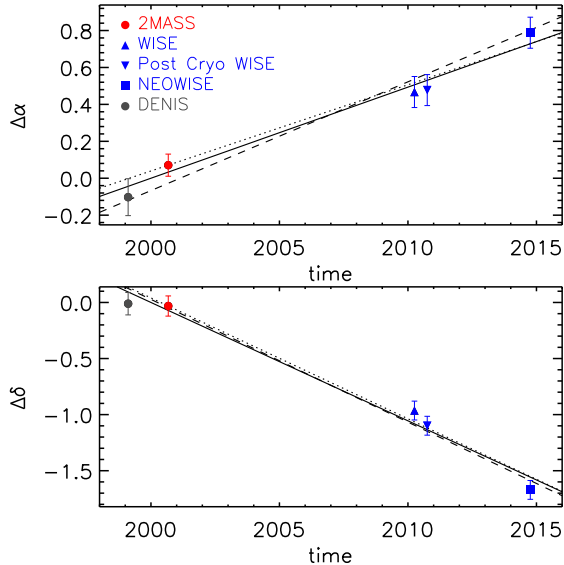}
\caption{\label{pm_fit} Our proper motion fit for \protect\secnameshort\enspace using data from various infrared surveys. The solid line is our proper motion fit, the dashed line is the proper motion for \protect\priname\enspace \protect\citep{Zacharias2013} shifted so it matches our proper motion at the midpoint of our dataset and the dotted line is the \protect\cite{Gagne2014} proper motion extrapolated from the 2MASS position.}
\end{center}
\end{figure}

\begin{figure*}
\begin{center}
\includegraphics[scale=0.5,angle=-90]{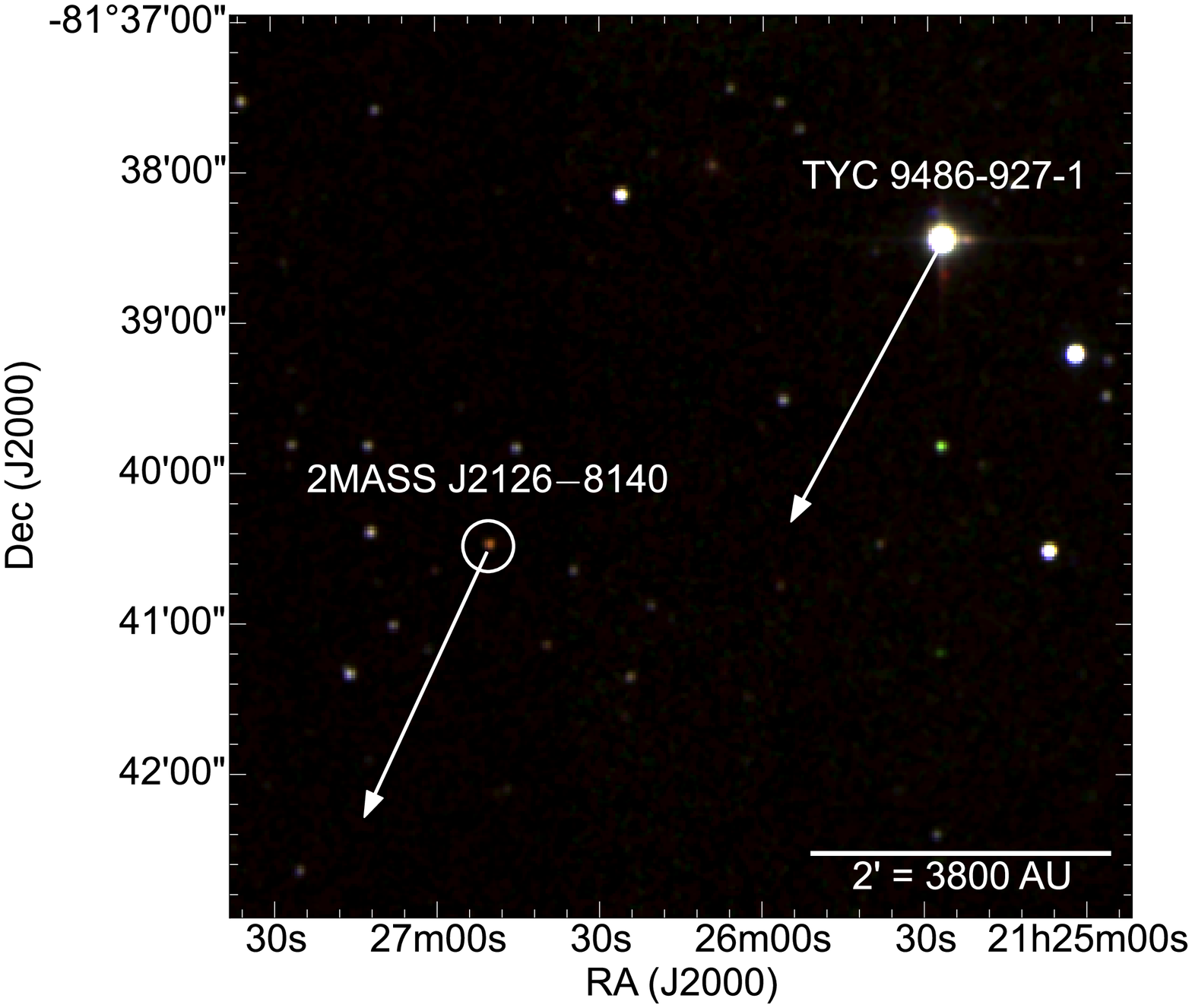}
\caption{A 6$\times$6\,arcminute {\it 2MASS} $JHK_s$ colour image of \protect\priname\enspace and \protect\secnameshort\enspace showing the direction of the proper motion vectors from \protect\cite{Zacharias2013} and this work respectively.  The length of the arrows shows the motion on the sky over 1000 years. The AU scale assumes the \protect\cite{Filippazzo2015} distance of 31.9\,pc.}
\label{pm_im}
\end{center}
\end{figure*}

Using 2MASS and WISE astrometry, \cite{Gagne2014} measured proper motions of $\mu_{\alpha}\cos\delta=46.7\pm1.3$\,mas/yr and $\mu_{\delta}=-107.8\pm10.4$\,mas/yr for \secnameshort.~This is deviant by 6$\sigma$ in the R.A direction from the UCAC4 measurements \citep{Zacharias2013} for \priname\enspace of $\mu_{\alpha}\cos\delta=58.9\pm1.5$\,mas/yr and $\mu_{\delta}=-109.4\pm1.0$\,mas/yr. Due to the very small uncertainty on \cite{Gagne2014}'s $\mu_{\alpha}\cos\delta$ measurement and the availability of newer datasets we recalculated the astrometric solution for \secnameshort\enspace using data from 2MASS \citep{Skrutskie2006}, the WISE All-Sky release\citep{Wright2010}, one epoch of WISE post-cryo data, one epoch of the reactivated NEO-WISE mission \citep{Mainzer2011} and the DENIS survey \citep{Epchtein1994}.  For each of the three WISE epochs we averaged the single exposure positional measurements to produce three datapoints. We assumed positional errors (84\,mas on both axes) from the quoted errors on the position for \secnameshort in the WISE All-Sky data release Source Catalogue and applied these to all three of our WISE datapoints \footnote{The WISE All-Sky Source Catalogue position is the average of multiple measurements at one of our epochs and thus the error on this averaged position will be representative of the error on our averaged position at each of our three WISE epochs.}. For 2MASS we used the quoted positions and positional error and for DENIS we used the approach of \cite{Luhman2013}, measuring the positional scatter on objects of similar brightness close to the target. This latter calculation yielded positional uncertainties of 100\,mas in both R.A. and Dec. which were applied to positions averaged from the different DENIS epochs. These measurements were combined in a least squares fit which resulted in proper motion measurements of $\mu_{\alpha}\cos\delta=49.3\pm9.7$\,mas/yr and $\mu_{\delta}=-105.5\pm6.6$\,mas/yr. These figures deviate by less than $1\sigma$ from the UCAC4 proper motion measurements for \priname\enspace from \cite{Zacharias2013}. Our proper motion fit along with those from \cite{Gagne2014} and the UCAC4 proper motion for \priname\enspace are shown in Figure~\ref{pm_fit}. The congruent proper motions are readily apparent on the plane of the sky in Figure~\ref{pm_im}.

\begin{table}
\caption{The radial velocity measurements derived from the low signal to noise Gemini/Phoenix spectrum of \protect\secnameshort\enspace by cross-correlating with different spectral templates. The three spectral template stars come from \protect\cite{Prato2002} and the model template from \citep{Allard2010}.}

\label{sec_rv}
\begin{center}
\scriptsize
\begin{tabular}{lrc}
\hline
Template&RV (km/s)&Cross-correlation power\\
\hline
1800\,K model&10.5$\pm$1.1&18\%\\
LHS 2065 (M9)&7.4$\pm$1.8&23\%\\
LHS 2351 (M7)&8.4$\pm$1.4&24\%\\
LHS 292 (M6)&8.1$\pm$1.4&23\%\\
GL 644C (M7)& 7.6$\pm$1.5&24\%\\
\hline
Adopted value&8.4$\pm$2.1\\
\hline
\normalsize

\end{tabular}
\end{center}
\end{table}
\subsection{The radial velocity of 2MASS~J21265040$-$8140293}
 \secnameshort~was observed with the Phoenix spectrograph \citep{Hinkle2003} mounted to the Gemini-South telescope on UT 2009 October 29 (Programme GS-2009B-C-2, PI K. Cruz). The observations consisted of two AB pairs with each exposure lasting 1800s. The data was obtained in the H-band with the 0.34 arcsecond slit which provides a resolving power of approximately 50,000. Along with the science data, flat lamp and dark calibration exposures were obtained on the same night. These data were downloaded from the Gemini Archive and reduced using a series of custom IDL routines. We corrected for bad pixels then flat-fielded and dark subtracted the science frames using a median master flat and dark frame. We attempted to extract the one dimensional spectrum from both sky-subtracted AB pairs, but the trace was only detected by our software in one pair. OH night sky lines were used to solve the dispersion solution and establish a wavelength scale. The final extracted spectrum covered 1.5512 - 1.5577 microns and had a SNR$\sim$5. Prior to cross correlation with model and observed templates, the spectrum was flattened by dividing by a 4th order polynomial fit to the continuum to the continuum, flux normalized, and corrected for the barycentric velocity. 

To measure the RV of 2MJ2126, we cross-correlated the spectrum with a 1800\,K, log(g)=4.0, solar metallicity BT Settl \citep{Allard2010} model spectrum (to match the parameters from \citealt{Manjavacas2014}) and observed M6, M7, and M9 Keck/NIRSPEC template spectra from \cite{Prato2002}. Our analyses provide consistent RVs using each template, all four measurements are listed in Table~\ref{sec_rv}. All measurements are in reasonable agreement with \priname's\enspace RV (10.0$\pm$1.0\,km/s, see Section~\ref{tyc_char}). We adopt a range of 8.4$\pm$2.1\,km/s for \secnameshort's\enspace radial velocity based on our measurements. We note that this radial velocity estimate comes from a very low signal to noise spectrum. However we include this measurement to demonstrate that we have analysed the available archive data for \secnameshort\enspace and can find no data which suggests that it is not in a bound system with \priname.

\subsection{Characterisation of TYC 9486-927-1}
\label{tyc_char}
TYC 9486-927-1 was classified as an active M1 by \cite{Torres2006} who also detected a Lithium 6708\AA\, feature with an equivalent width of 104m\AA.\, We observed TYC 9486-927-1 with FEROS (\citealt{Kaufer1999}; R=48,000, 3600--9200\AA)\enspace on 2012 October 7 using all the standard settings, reductions, and RV analyses detailed in sections 2.3.7 and 3.11 of \cite{Bowler2015}. We subsequently observed the star on 2015 August 26 and 2015 October 27 with the WiFeS instrument on the ANU 2.3\,m telescope at Siding Spring using the R7000 grating (\citealt{Dopita2007}; R=7000, 5250--7000\AA).\enspace The WiFeS instrument set up, data reduction and analysis, including the derivation of  line widths and radial velocities, was the same as that described in \cite{Murphy2014}. Our FEROS spectrum showed emission in H$\alpha$, H$\beta$, H$\gamma$ and H$\delta$ and yielded measurements of $v_{rad}=10.0\pm1.0$\,km/s,$EW_{Li}=85\pm$15m\AA~ and $v\sin i=40.0\pm2.0$\,km/s, while the WiFeS observations measured $v_{rad}=10.7\pm1.0$\,km/s and 9.7$\pm$1.0\,km/s respectively with $EW_{Li}=90\pm$10m\AA in both observations. Table~\ref{TYC_spec} shows the spectroscopic properties of \priname\enspace  from our observations and the literature. Notably, the star exhibits no  RV variations outside of uncertainties on  the scale of months to years, strongly suggesting it is not a close to equal mass spectroscopic binary system. We adopt a radial velocity of 10.0$\pm$1.0\,km/s for the star, in agreement with the FEROS and WiFeS data, and lower precision measurements of 8.7+/-4.6 km/s \citep{Torres2006} and 11.9$\pm$3.0\,km/s \citep{Malo2014}. To further examine the possible binary nature of \priname,\enspace we performed 2D cross correlations on our FEROS spectrum with a variety of primary/secondary template mass ratio combinations. None of these tests yielded a reliable cross-correlation function with power larger than in the case of a single star.

\begin{figure}
\begin{center}
\includegraphics[scale=0.5,angle=-90]{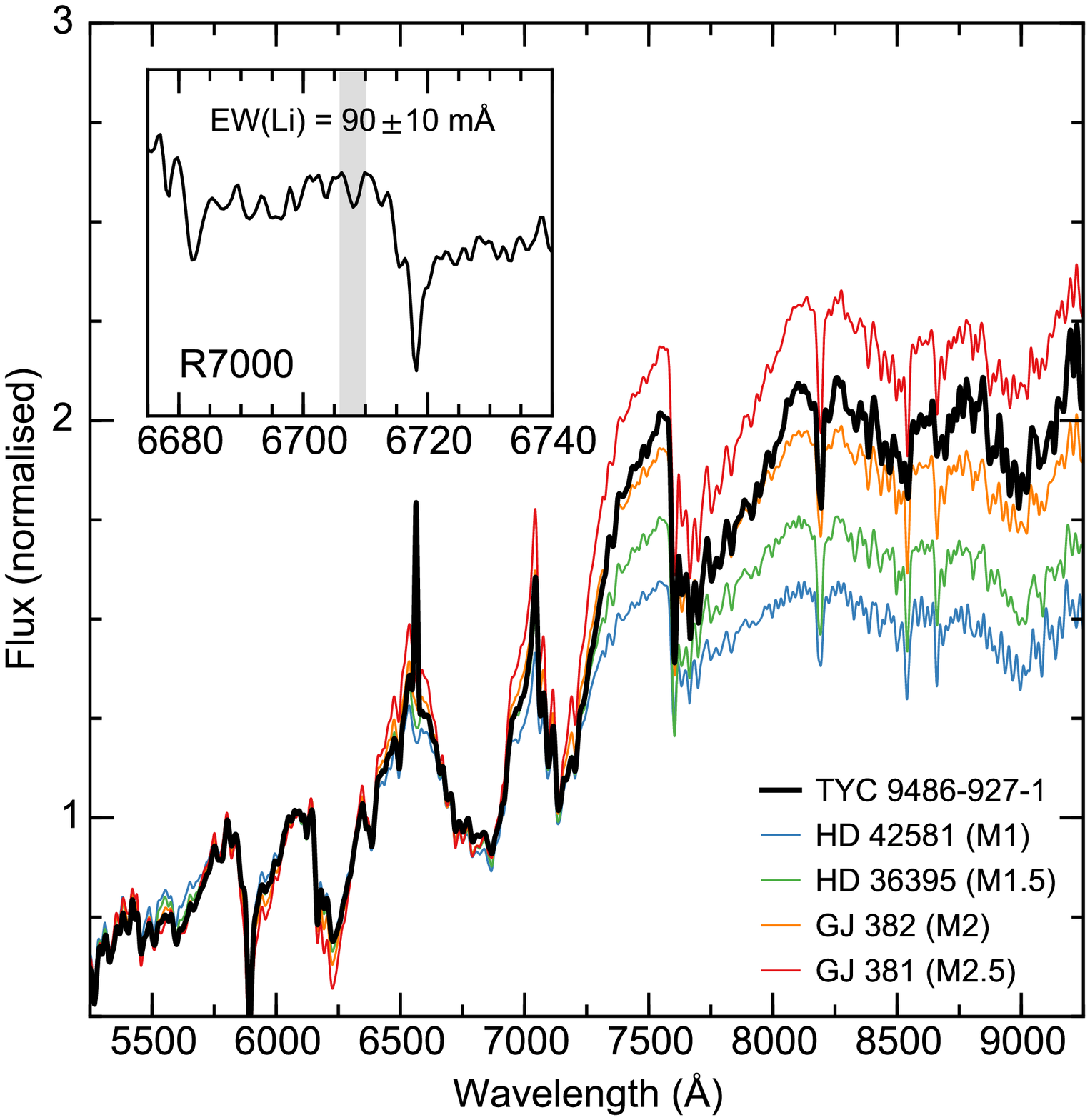}
\caption{.WiFeS/R3000 spectrum of TYC 9486-927-1 (thick line), compared to spectral type standards observed the same night. All spectra have been normalised at 6100A and smoothed with a 5-px Gaussian prior to plotting.  Based on this comparison the spectral type of TYC 9486-927-1 is M2-M2.5.  The inset shows the weak (EW=90$\pm$10 m\AA)\enspace and broad Li I 6708\AA\enspace line from a higher resolution WiFeS R7000 observation.}
\label{wifes_spec}
\end{center}
\end{figure}

To garner an improved spectral type, we made an additional observation of \priname\enspace with WiFeS on the 2015 November 28 with the lower resolution R3000 grating  (R=3000, 5200-9800A). We visually compared the flux calibrated and telluric corrected spectrum to spectral type standards from the lists of E. Mamajek\footnote{\url{http://www.pas.rochester.edu/~emamajek/spt/}} observed that night with the same instrument settings. Figure~\ref{wifes_spec} shows that \priname\enspace has a spectral type between M2 (GJ 382) and M2.5 (GJ 381), inconsistent with the M1 spectral type reported by \cite{Torres2006}. To compliment the visual comparison we also measured several molecular spectral type indices recently calibrated by \cite{Lepine2013}. These are listed in Table~\ref{TYC_ind} compared to measurements made from a low-resolution spectrum of TYC 9486-927-1 by \cite{Gaidos2014}, who assigned a spectral type of M3 from visual inspection. Based on all available spectroscopic  observations we assign a spectral type of M2.0$\pm$0.5 for \priname.  This also agrees with photometric spectral types obtained from $V-J$ (M2.3, \citealt{Lepine2013}) and  $V-K_s$ (M1.8, \citealt{Pecaut2013}) colours.

 \cite{Elliott2015} imaged TYC 9486-927-1 with the VLT/NACO AO imager and found no companion, despite being able to detect an equal mass companion down to a projected separation of 3\,AU (at an assumed photometric distance of 36.3\,pc distance or 4.2\,AU adjusting that photometric distance for equal-mass binarity). These observations and our multi-epoch radial velocity data suggest that TYC 9486 is a single, rapidly rotating star rather than a spectroscopic or tight, visual binary. However, it is still possible that TYC~9486-927-1 is an equal mass binary with a face-on orbit and close separation.

\begin{table}
\caption{Properties derived from multi-epoch spectroscopy for \protect\priname\enspace from both the literature and our work. Note the consistency of the radial velocity over long periods and the variability in the H$\alpha$ EW due to chromospheric activity.. The last line refers to our lower resolution R3000 observation where the Lithium 6708\AA\,feature was not resolved.}

\label{TYC_spec}
\begin{center}
\scriptsize
\begin{tabular}{lllcccc}
\hline
Source&Date (UT)&SpT&RV(km/s)&EW Li (m\AA)&EW H$\alpha$ (\AA)&$v\sin i$(km/s)\\
\hline
\protect\cite{Torres2006}&2001-09-08&M1e&8.7$\pm$4.6&104&-5.6&43.5$\pm$1.2\\
\protect\cite{Malo2014}&2010-05-25&\ldots&11.9$\pm$3.0&\ldots&\ldots&44.8$\pm$4.2\\
\protect\cite{Gaidos2014}&\ldots&M3&\ldots&\ldots&-3.9&\ldots\\
This work, FEROS&2012-10-07&\ldots&10.0$\pm$1.0&85$\pm$15&\ldots&40.0$\pm$2.0\\
This work, WiFeS&2015-08-26&\ldots&10.7$\pm$1.0&90$\pm$10&-5.7$\pm$0.2&\ldots\\
&2015-10-27&\ldots&9.7$\pm$1.0&90$\pm$10&-9.5$\pm$0.2&\ldots\\
&2015-11-28&M2$\pm$0.5&\ldots&\ldots&-6.0$\pm$0.5&\ldots\\
\hline
\normalsize

\end{tabular}
\end{center}
\end{table}
\begin{table}
\caption{Spectral indices in the \protect\cite{Lepine2013} format and the estimated spectral type from each value (in parenthesis). Our visual comparison is relative to the spectral standards of Mamajek\protect\footnote{\protect\url{http://www.pas.rochester.edu/~emamajek/spt/}}. }

\label{TYC_ind}
\begin{center}
\scriptsize
\begin{tabular}{lcccccccc}
\hline
Source&CaH2&CaH3&TiO5&VO1&VO2&ColorM&Visual&Final\\
\hline
\protect\cite{Gaidos2014}&0.613(M1.2)&0.793(M1.6)&0.697(M1.2)&\ldots&\ldots&\ldots&M3&\\
This work&0.574(M1.7)&0.769(M2.1)&0.591(M2.2)&0.940(M2.7)&0.886(M2.4)&1.575(M3.1)&M2&M2\\
\hline
\normalsize

\end{tabular}
\end{center}
\end{table}

\subsubsection{The age of TYC 9486-927-1}
\begin{figure}
\begin{center}
\includegraphics[scale=0.5]{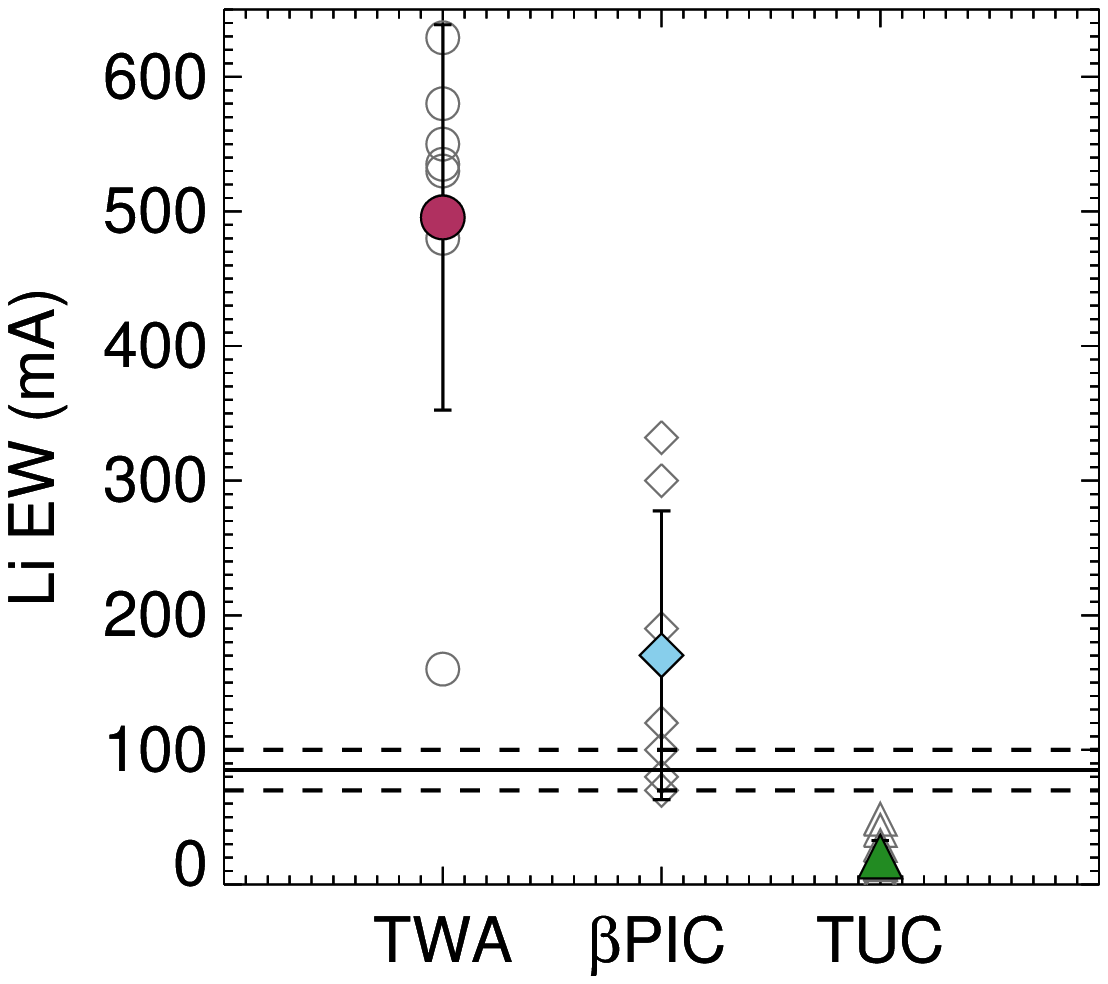}
\caption{\label{Li_plot} The lithium 6708\AA\, absorption for \protect\priname (solid line with dashed line error bars) plotted against M1 and M2 members \protect\citep{DaSilva2009} of the TWA (10$\pm$3\,Myr), $\beta$~Pic (24$\pm$3\,Myr) and TucHor (45$\pm$4\,Myr) associations (all ages from \protect\citealt{Bell2015}) with additional TucHor members from \protect\cite{Kraus2014a}. The solid data points with error bars show the mean and standard deviation of each association. Clearly \protect\priname has a lithium absorption strength between members of TWA and TucHor and in agreement with $\beta$~Pic.}
\end{center}
\end{figure}

\priname\enspace has rapid rotation and coronal and chromospheric activity suggestive of a young age. We measured a weak lithium 6708\AA\, absorption feature in both our FEROS (85$\pm$15m\AA)\enspace and WiFeS spectra (90$\pm$10m\AA, see Figure~\ref{wifes_spec} inset)\enspace consistent with the 104m\AA\enspace measurement of \cite{Torres2006}. The detection of lithium is an important age diagnostic for early M-dwarfs, as the element is typically depleted in the photospheres of such stars on time scales of <40 Myr (e.g. \citealt{Mentuch2008}). To illustrate this, in Figure~\ref{Li_plot} we show the Li I 6708\AA\enspace EWs of M1 and M2 type stars from the SACY sample \citep{DaSilva2009} for the TW Hydrae (TWA; 10$\pm$3\,Myr), $\beta$~Pic (24$\pm$3\,Myr) and TucHor (45$\pm$4\,Myr) associations with additional TucHor members from \cite{Kraus2014a} (all ages from \citealt{Bell2015}) against the EW for \priname\enspace. It is clear that the \priname\enspace has stronger lithium absorption than stars of similar spectral type in TucHor, weaker absorption than TWA members but in reasonable agreement with $\beta$~Pic members. Based on this comparison we conclude that \priname\enspace is older than TWA and likely of similar age or younger than TucHor. Thus, our Li analysis suggests an age comparable to the $\beta$ Pic moving group, but we note that Li depletion in low-mass stars can be affected by initial conditions (rotation, episodic accretion) and we therefore adopt a conservative age range of 10--45 Myr.

\begin{table}
\caption{Summary of the properties of both \protect\priname\enspace and \protect\secnameshort.}

\label{sys_sum}
\begin{center}
\footnotesize
\begin{tabular}{lcc}
\hline
&\protect\priname&\protect\secnameshort\\
\hline
Position (J2000)&21 25 27.52 $-$81 38 27.8$^a$&21 26 50.40 $-$81 40 29.3$^a$\\
$\mu_{\alpha}\cos\delta$ (mas/yr)&58.9$\pm$1.5$^{b}$&49.3$\pm$9.7$^{c}$\\
$\mu_{\delta}$ (mas/yr)&$-$109.4$\pm$1.0$^{b}$&$-$105.5$\pm$6.6$^{c}$\\
$v_{rad}$ (km/s)&10.0$\pm$1.0$^{c}$&8.4$\pm$2.1$^{c*}$\\
$J$ (mag.)&8.244$\pm$0.03$^{g}$&15.54$\pm$0.06$^{g}$\\
$H$ (mag.)&7.563$\pm$0.027$^{g}$&14.40$\pm$0.05$^{g}$\\
$K_s$ (mag.)&7.34$\pm$0.04$^{g}$&13.55$\pm$0.04$^{g}$\\
$W1$ (mag.)&7.241$\pm$0.032$^{h}$&12.910$\pm$0.024$^{h}$\\
$W2$ (mag.)&7.151$\pm$0.021$^{h}$&12.472$\pm$0.023$^{h}$\\
$W3$ (mag.)&7.039$\pm$0.016$^{h}$&11.885$\pm$0.161$^{h}$\\
$W4$ (mag.)&6.953$\pm$0.061$^{h}$&$>$9.357$^{h}$\\
Age&10--45\,Myr$^{c}$&10--150\,Myr$^f$\\
Mass&$\sim$0.4M$_{\odot}$\,$^{c}$&11.6--15.0M$_J$\,$^{c}$\\
Distance (pc)&20.5-29.0$^{c}$&31.9$^{+2.9}_{-2.4}$$^d$\\
Spectral Type&M2$^c$&L3$\gamma$$^{e}$\\
Separation&\multicolumn{2}{c}{217\,\protect\arcsec}\\
&\multicolumn{2}{c}{$\sim$6900\,AU}\\
\hline
\multicolumn{2}{l}{$^{a}$ 2MASS position, epoch 2000.644 \protect\cite{Skrutskie2006}}\\
\multicolumn{2}{l}{$^{b}$ \protect\cite{Zacharias2013}}\\
\multicolumn{2}{l}{$^{c}$ This work}\\
\multicolumn{2}{l}{$^{d}$ Trigonometric parallax mentioned in \protect\cite{Filippazzo2015} citing Faherty et al. in prep.}\\
\multicolumn{2}{l}{$^{e}$ \protect\cite{Faherty2013}}\\
\multicolumn{2}{l}{$^{f}$ \protect\cite{Filippazzo2015}}\\
\multicolumn{2}{l}{$^{g}$ \protect\cite{Skrutskie2006}}\\
\multicolumn{2}{l}{$^{h}$ \protect\cite{Wright2010}}\\
\multicolumn{2}{l}{$^{*}$ Measurement from a very low signal to noise spectrum.}\\
\normalsize

\end{tabular}
\end{center}
\end{table}

\subsection{Photometric distances and moving group membership}
\priname\enspace lacks a trigonometric parallax measurement and thus any determination of its kinematics (and hence moving group membership) requires photometric distance estimates. To estimate absolute near-IR magnitudes for \priname\enspace we used our measured spectral type of M2. We then derived an effective temperature of 3490\,K for \priname\enspace using the 5--30\,Myr young star $T_{eff}$ scale of \cite{Pecaut2013} and applied this to the evolutionary models of \cite{Baraffe2015} at four ages (10, 20, 30 and 40\,Myr) to estimate the absolute magnitudes. \priname's $J$, $H$ and $K_s$ 2MASS photometry were then compared to the these absolute magnitudes to calculate distances, neglecting the likely negligible extinction. We took the mean of these estimates as the adopted distance for \priname\enspace for each age (see Table~\ref{TYC_dist}). Binarity would change the photometric distances although our multi-epoch RV measurements and the high resolution imaging of \cite{Elliott2015} show no evidence of a close companion to \priname. 

To compare to the trigonometric parallax quoted in \cite{Filippazzo2015} for \secnameshort\enspace we used the young L dwarf photometric distance relations of \cite{Gagne2015a}. Following a similar process to that described above but adopting the scatter on the relations as the error on our distances. As the \cite{Gagne2015a} relations cover a wide range of ages (up to 125\,Myr) they also cover a wide range of luminosities for each spectral type due to young objects having inflated radii. Hence the photometric distances do not deviate randomly across bands but will be correlated. Thus we do not adopt a weighted mean distance but take the distance from the band with the lowest scatter ($W2$, d= 26.7$^{+5.7}_{-4.7}$\,pc). This distance, and those for TYC 9486-927-1 using the 10 and 20\,Myr \cite{Baraffe2015} models agree well with the trigonometric parallax presented by \cite{Filippazzo2015}.

\begin{figure}
\begin{center}
\includegraphics[scale=0.5,angle=-90]{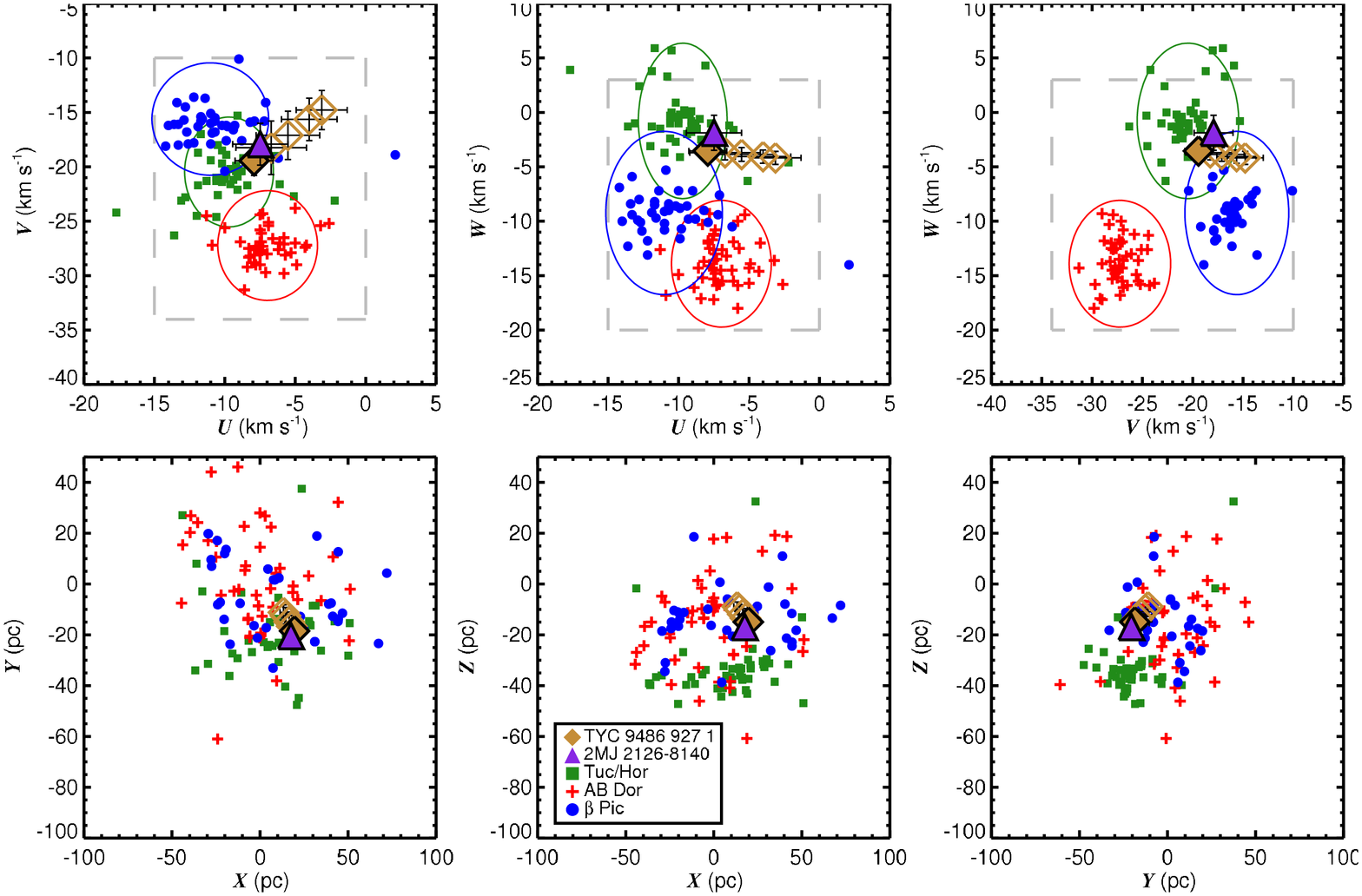}
\caption{\label{kin_plot} Projections of Galactic velocities and positions for members of the Tucana/Horologium association, the AB Doradus moving group, and the $\beta$ Pictoris moving group \protect\citep{Malo2013} along with 3$\sigma$ velocity ellipsoids \protect\citep{Gagne2014}. Top panels: We show for comparison the UVW velocities of TYC9486 using the measured proper motions, RV, and a range photometric distances for estimated ages 10--40 Myr (open data points left to right) and the secondary trigonometric parallax from \protect\cite{Filippazzo2015} (solid symbol). The velocities of \protect\secnameshort\enspace are calculated using our proper motion, the RV from the low signal to noise Phoenix spectrum, and the parallax distance from \protect\cite{Filippazzo2015}. The velocities of both compoents are approximately consistent with both TucHor and $\beta$Pic for distances around 30 pc. Bottom panels: Same as the top but for XYZ positions the positions of the primary have been offset by +2\,pc for clarity. The 10--40\,Myr photometric distance estimates for TYC9486 (bottom to top) and the trigonometric distance for 2MJ2126 are consistent with the XYZ distribution of $\beta$Pic but are TucHor outliers in Z.}
\end{center}
\end{figure}

To test the membership of TYC 9486-927-1 in several well known moving groups in the solar neighbourhood, we imputed our radial velocity (10$\pm$1\,km/s), the UCAC4 proper motions and the positions, along with the distance estimates for each age, into the BANYAN\,II young moving group membership probability estimation tool \citep{Malo2013,Gagne2014}. We assumed a 20\% error on our photometric distance estimates and that the objects were younger than 1\,Gyr. For \secnameshort\enspace we used our proper motion, and both the photometric distance estimate and the literature trigonometric parallax. The results of these calculations are shown in Table~\ref{TYC_dist}. They suggest that the system is unlikely to be a TucHor member. Over the range of estimated photometric distances in Table~\ref{TYC_dist} , BANYAN II provides probabilities of $\beta$ Pic membership ranging from about 4.9 to 74\%. The membership probability for \secnameshort\enspace is on the lower end of this range when we allow the radial velocity to be unconstrained. To further investigate the potential moving group membership we plotted the Galactic $U,V,W$ velocities and $X,Y,Z$ positions for \priname\enspace and \secnameshort\enspace (Figure~\ref{kin_plot}). We find that the reason BANYAN discounts TucHor membership is due to the system being a significant outlier in the $Z$ coordinate. While $\beta$~Pic remains the most likely moving group (both in kinematics and in age) for this system to be associated with, a difference in the $U$ velocity precludes us from claiming this is a bona-fide $\beta$~Pic member.

\begin{table*}
\begin{minipage}{170mm}
\caption{Photometric distances and moving group membership probabilities (TH - Tuc Hor, BP $\beta$ Pic, YF young field) for \protect\priname\enspace and \protect\secnameshort.\enspace We assume errors of $\sim$20\% for the photometric distance estimates for \priname. Membership probabilities come from the BANYAN~II online tool \protect\citep{Malo2013,Gagne2014} and use the proper motions from \protect\cite{Zacharias2013} for \protect\priname\enspace and our own calculation for \protect\secnameshort. Calculations using measured radial velocities for \protect\priname\enspace (10.0$\pm$1.0\,km/s) and \secnameshort\enspace (8.4$\pm$2.1\,km/s) are marked accordingly. We note again this latter radial velocity is derived from a low signal to noise spectrum. See Table~\protect\ref{sys_sum} for a full list of derived parameters. The final line for each object uses the trigonometric parallax mentioned in \protect\cite{Filippazzo2015}.}
\label{TYC_dist}
\begin{center}
\scriptsize
\begin{tabular}{cclccccccclll}
\hline
Object&SpT&Age&RV (km/s)&$d_J$&$d_H$&$d_K$&$d_{W1}$&$d_{W2}$&$d_{adopted}$&$p_{TH}$&$p_{BP}$&$p_{YF}$\\
\hline
%TYC 9486-927-1&M1&20\,Myr&31.6&31.5&31.5&\ldots&\ldots&31.5&1.0\%&32.5\%&64.5\%\\
%&M1&30\,Myr&28.1&27.8&27.9&\ldots&\ldots&28.0&0.0\%&66.0\%&33.7\%\\
%&M1&40\,Myr&25.8&25.6&25.6&\ldots&\ldots&25.7&0.0\%&75.4\%&24.5\%\\
TYC 9486-927-1&M2&10\,Myr&10.0$\pm$1.0&29.1$\pm$5.8&28.9$\pm$5.8&29.0$\pm$5.8&\ldots&\ldots&29.0$\pm$5.8&0.0\%&59.1\%&40.2\%\\
&M2&20\,Myr&10.0$\pm$1.0&26.4$\pm$5.3&26.0$\pm$5.2&26.1$\pm$5.2&\ldots&\ldots&26.2$\pm$5.2&0.0\%&74.0\%&25.8\%\\
&M2&30\,Myr&10.0$\pm$1.0&22.8$\pm$4.6&22.3$\pm$4.5&22.5$\pm$4.5&\ldots&\ldots&22.6$\pm$4.5&0.0\%&71.8\%&28.18\%\\
&M2&40\,Myr&10.0$\pm$1.0&20.7$\pm$4.1&20.3$\pm$4.1&20.5$\pm$4.1&\ldots&\ldots&20.5$\pm$4.1&0.0\%&55.6\%&44.4\%\\
&\ldots&\ldots&10.0$\pm$1.0&\ldots&\ldots&\ldots&\ldots&\ldots&31.9$^{+2.9}_{-2.4}$&1.6\%&9.5\%&88.3\%\\
2MASS~J2126$-$8140&L3&$<$125\,Myr&\ldots&26.9$^{+19.0}_{-11.1}$&26.2$^{+10.2}_{-7.4}$&24.9$^{+7.3}_{-5.7}$&26.7$^{+8.0}_{6.2}$&26.7$^{+5.7}_{-4.7}$&26.7$^{+5.7}_{-4.7}$&0.0\%&29.8\%&68.8\%\\
&\ldots&\ldots&\ldots&\ldots&\ldots&\ldots&\ldots&\ldots&31.9$^{+2.9}_{-2.4}$&0.3\%&4.9\%&91.7\%\\
&L3&$<$125\,Myr&8.4$\pm$2.1&26.9$^{+19.0}_{-11.1}$&26.2$^{+10.2}_{-7.4}$&24.9$^{+7.3}_{-5.7}$&26.7$^{+8.0}_{-6.2}$&26.7$^{+5.7}_{-4.7}$&26.7$^{+5.7}_{-4.7}$&0.0\%&69.4\%&30.6\%\\
&\ldots&\ldots&8.4$\pm$2.1&\ldots&\ldots&\ldots&\ldots&\ldots&31.9$^{+2.9}_{-2.4}$&1.0\%&34.8\%&64.0\%\\

\hline
\normalsize

\end{tabular}
\end{center}
\end{minipage}
\end{table*}
%The photometric distances and membership probabilities for each age and spectral type are shown in  Note that the previous suggestion that \priname\enspace is a member of TucHor is not supported by the membership probabilities with membership of $\beta$ Pic being far more likely. However if \priname\enspace were an equal mass binary, it may then still be a TucHor member. However our multi-epoch RV observations and the high resolution imaging of \cite{Elliott2015} do not support (but do not rule out altogether) the binary hypothesis. 

%As shown in Table~\ref{TYC_dist} the photometric distances of the two objects match well. \secnameshort\enspace has a lower probability of being a $\beta$ Pic member than \priname\enspace but this is due to the radial velocity of \secnameshort\enspace not being known. 

\subsection{Probability of chance alignment}
\begin{figure}
\begin{center}
\begin{tabular}{cc}

\includegraphics[scale=0.5]{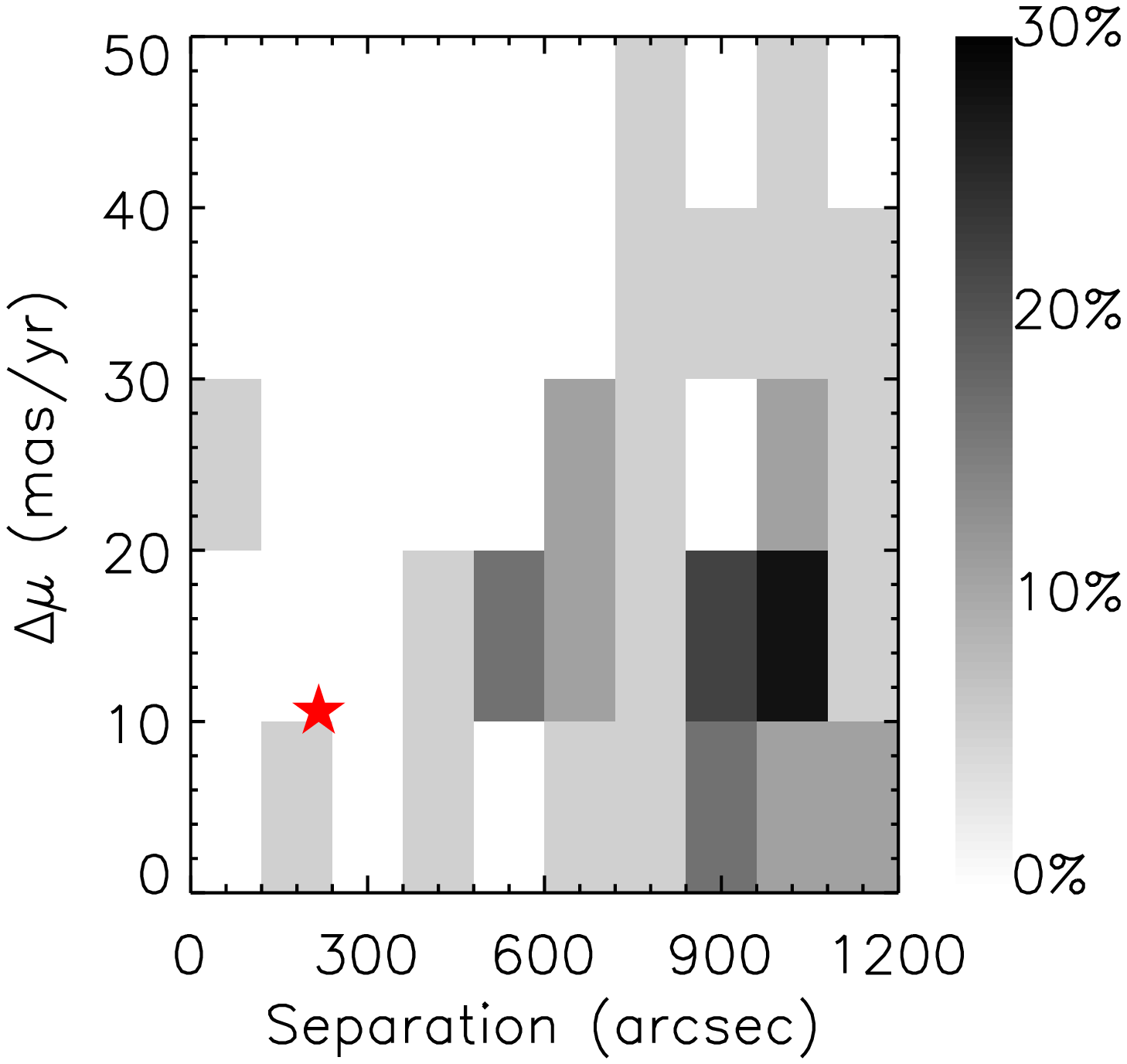}&
\includegraphics[scale=0.5]{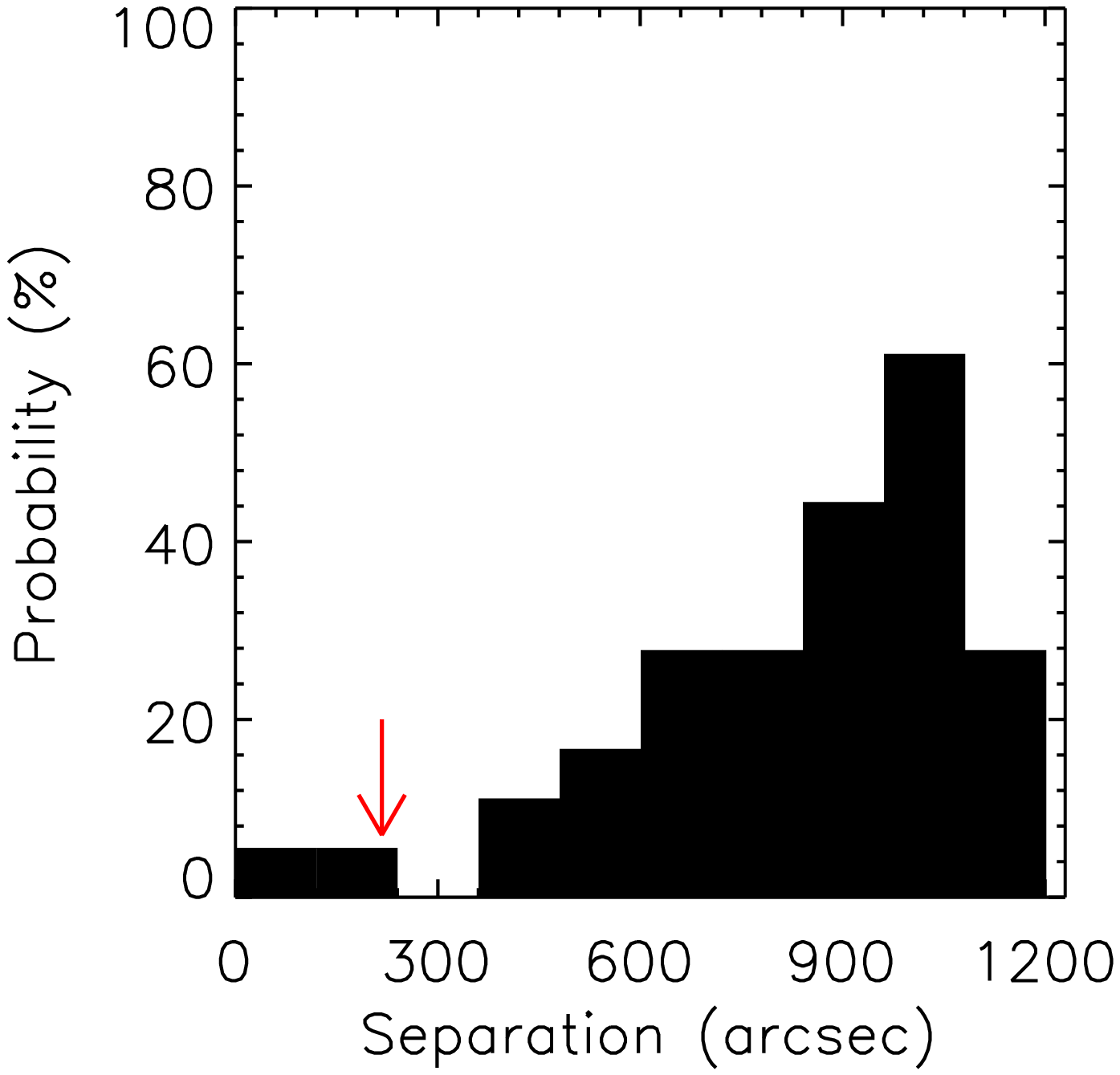}\\
\includegraphics[scale=0.5]{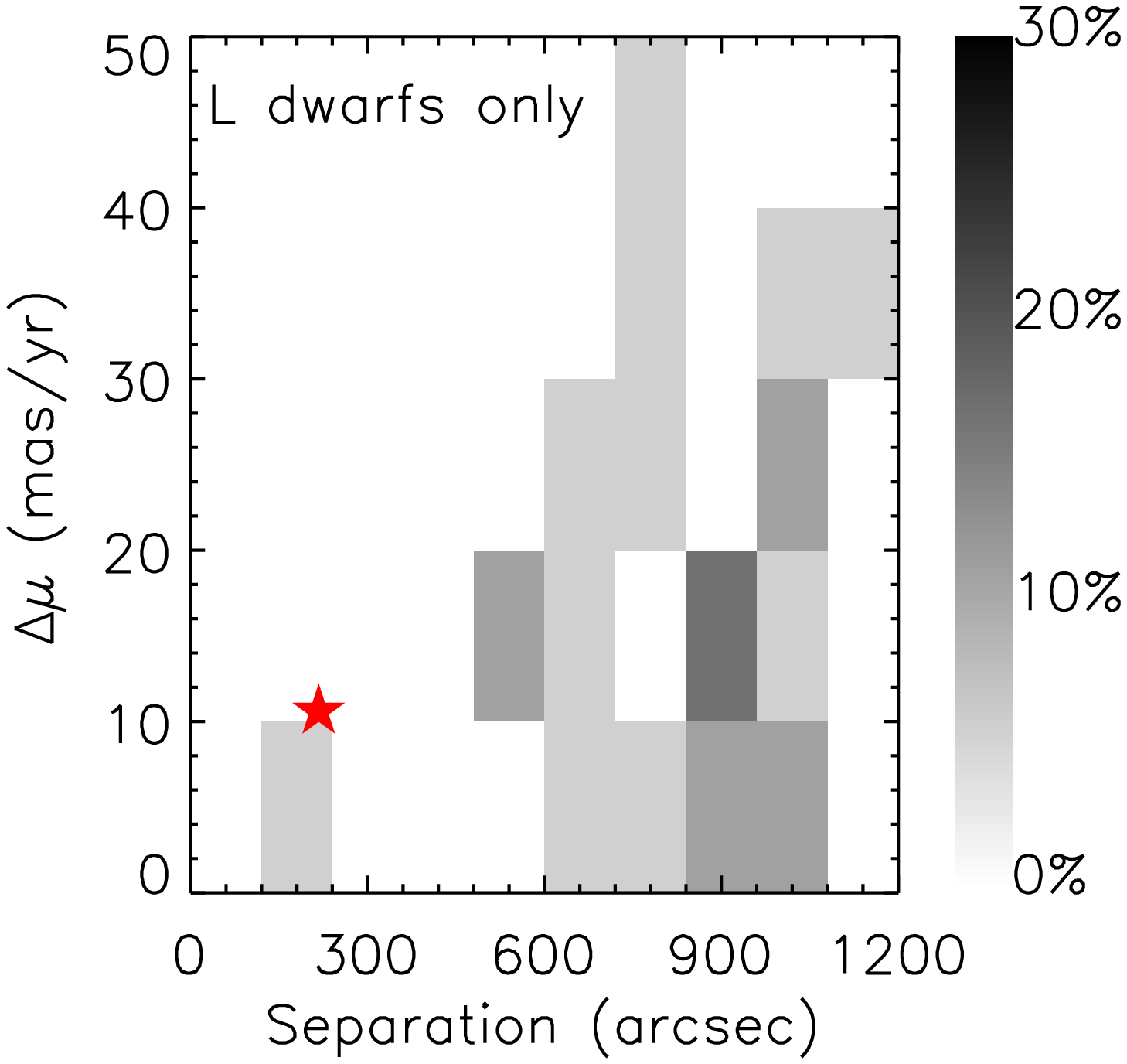}&
\includegraphics[scale=0.5]{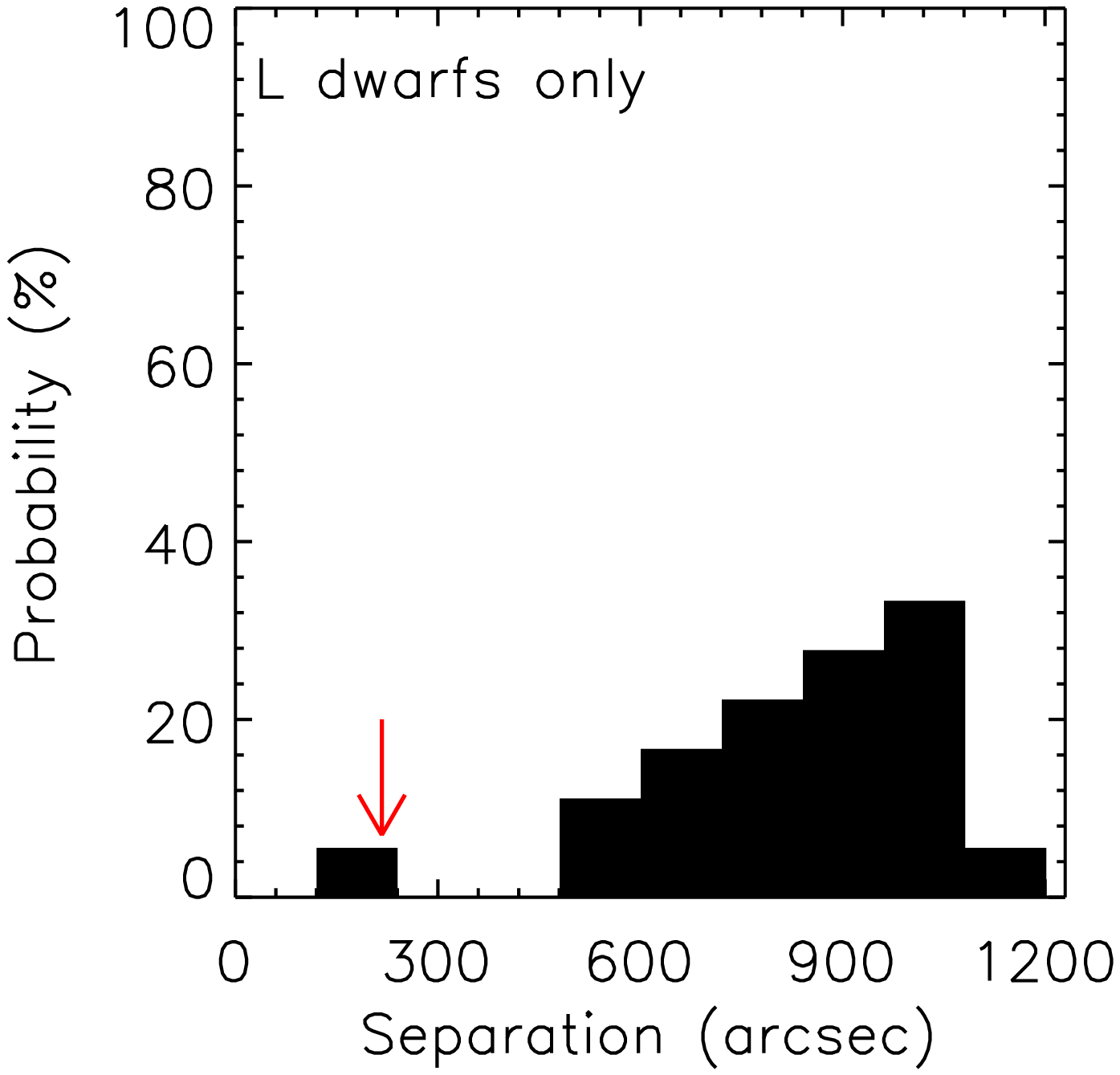}\\
\includegraphics[scale=0.5]{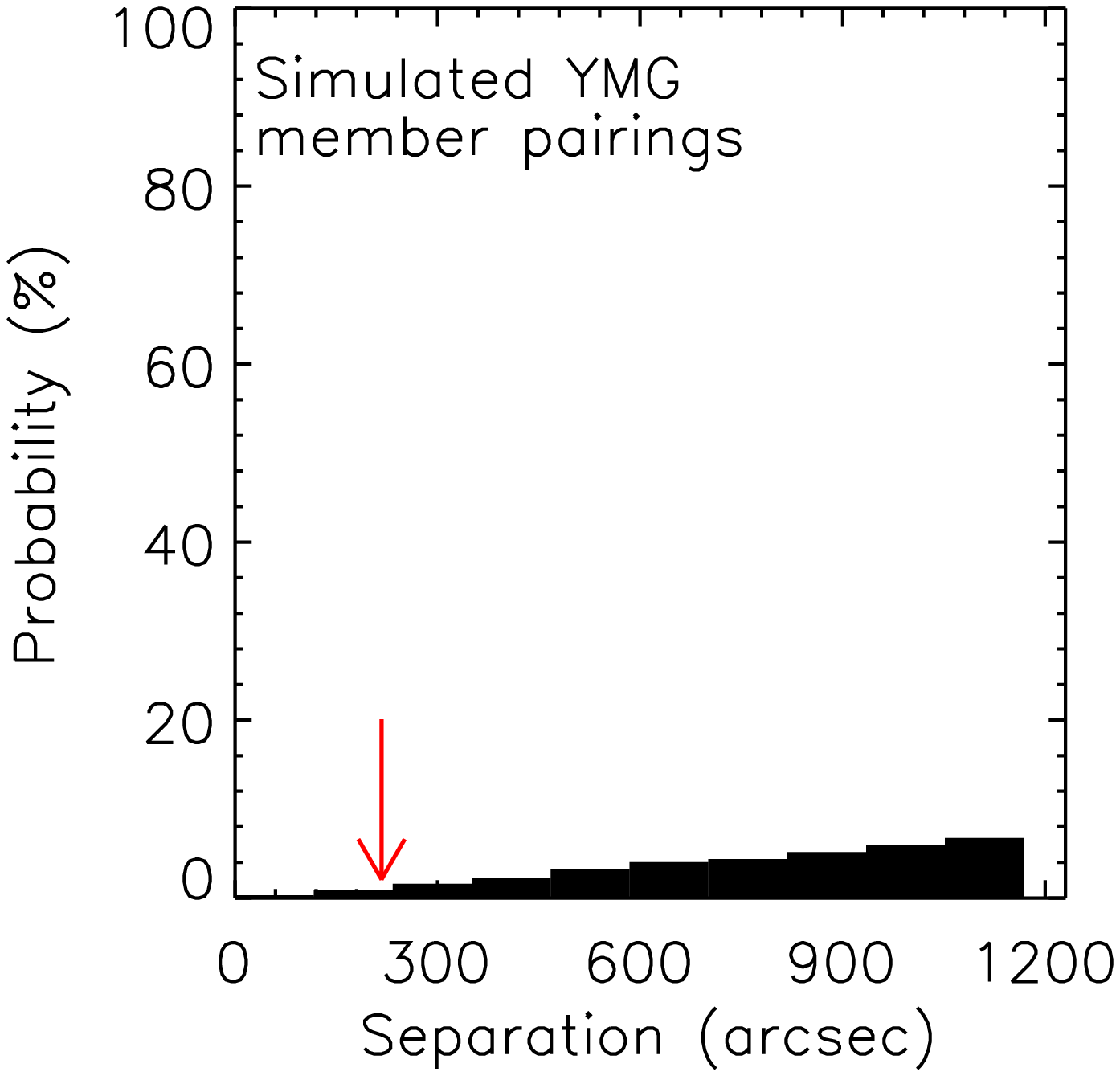}
\end{tabular}
\caption{\label{prob_lepine} {\bf Top two rows:}The results of our chance alignment using a modification of the offset position method of \protect\cite{Lepine2007}. The top row includes matches with all objects in \protect\cite{Gagne2015}'s catalogue while the middle row only includes matches with objects with $J-K_s>1.2$. {\bf Bottom row:} The probability of finding two members of the same moving group at a particular separation on the sky from each other and with distances which agree within 10\,pc. Clearly it is unlikely that our proposed pair (marked by a star or a red arrow) is a chance alignment.}
\end{center}
\end{figure}

While it appears that \priname\enspace and \secnameshort\enspace have matching proper motions and distances and both show signs of youth it is possible that they are a chance alignment of unrelated young objects. To determine the likelihood of this, we modified the method of \cite{Lepine2007}. We first constructed a list of known and candidate young stars in nearby young moving groups from \cite{Torres2008}, \cite{Shkolnik2009}, \cite{Schlieder2012}, \cite{Kraus2014a} and \cite{Malo2014}. We then offset the positions of these stars by two degrees and searched for common proper motion companions in the 2MASS-WISE proper motion catalogue of \cite{Gagne2015} around these offset positions. This should result in only chance alignments of unrelated objects. We ran this process 18 times, on each occasion offsetting the positions of our input sample by 2 degrees but changing the direction of the offset by one-ninth of a radian each time. In this way we were able to sample a much larger area for chance alignments and thus reduce statistical noise. The results are shown in Figure~\ref{prob_lepine} and clearly show a low probability ($\sim5\%$) of chance alignment considering all objects in Gagne's proper motion sample or only those with L dwarf-like colours ($J-K_s>1.2$). There is also the possibility of the chance alignment between two unbound objects in the same moving group. Hence we carried out a simulation to see how often two members of one group would fall close to each other on the sky and have photometric distance estimates within 10\,pc of each other. To accomplish this we generated a random realisation of each of the eight young moving groups described in \cite{Gagne2014} using the parameters provided by that work. We then ran this simulation 50,000 times and determined that there is $<$1\% chance that two objects in the same moving group would lie as close together on the sky as \priname\enspace and \secnameshort\enspace (see Figure~\ref{prob_lepine} lowest panel). We assumed that all chance alignments between members of one moving group would have matching proper motions due to both components having the bulk space velocity of the moving group. Note we did not consider the number of pairings between members of different moving groups (i.e. the number of chance alignments between AB~Dor and TWA members). Even accounting for a factor of 2 or 3 missing members in the groups (especially at the lowest masses), it is clear that TYC 9486-927-1 and 2MASS J2126-8140 are unlikely to be chance alignments inside the same group.

\section{Physical properties of 2MASS~J21265040$-$8140293}
In order to estimate the mass of \secnameshort,~we used a Monte Carlo simulation. We drew temperatures from a flat distribution from the 1700--1900\,K range quoted by \cite{Manjavacas2014} and ages from our flat 10--45\,Myr range. We then compared with the COND \citep{Baraffe2003} and \cite{Saumon2008} ($f_{sed}=2$) evolutionary models. The \cite{Baraffe2003} models yielded masses between 11.6\,$M_{Jup}$ and 14.7\,$M_{Jup}$ \footnote{The full range of masses derived, not a confidence interval.} and the \cite{Saumon2008} models preferred solutions in the 13.3--15\,$M_{Jup}$ range. Hence we adopt a mass range of 11.6--15\,$M_{Jup}$ range for \secnameshort\enspace placing it on the 13\,$M_{Jup}$ deuterium-burning dividing line between planets and brown dwarfs. A similar calculation drawing the age from a \citep{Bell2015} $\beta$ Pic Gaussian age distribution of 24$\pm$3\,Myr yields a mass range of 12--14$M_J$ . Such masses and ages make 2MASS J2126-8140 an interesting wide-orbit analogue to $\beta$~Pic~b \citep{Lagrange2010}, whose primary is a member of the eponymous moving group.. \cite{Morzinski2015} find that $\beta$~Pic~b has a mass of 12.7$\pm$0.3 $M_{Jup}$ and  $T_{eff}$ of 1708$\pm$23\,K whilst \cite{Bonnefoy2013} find a spectral type of L1$^{+1.0}_{-1.5}$, $T_{eff} = 1700\pm100$\,K, $\log g = 4.0 \pm 0.5$\,dex and mass of $10_{-2}^{+3} M_{Jup}$ from $T_{eff}$ and $9_{-2}^{+3} M_{Jup}$ from luminosity. These compare well with our derived age range, 11.6--15\,$M_{Jup}$ mass range and \cite{Manjavacas2014}'s $T_{eff}=1800\pm100$\,K and $\log g = 4.0\pm0.5$\,dex for \secnameshort. We note that $\beta$~Pic~b is commonly referred to as a planet and also has mass estimates straddling the deuterium burning limit. Our system is also a younger analogue of the AB Dor (149$^{+51}_{-19}$\,Myr; \citealt{Bell2015}) member GU~Psc A/b system \citep{Naud2014} with both components having similar masses.

The range of photometric distances for the system gives projected separations of 4450-5700\,AU and the trigonometric parallax of the secondary a separation of 6900\.AU, wider than any star-planet system in the exoplanet.eu database (\url{http://exoplanet.eu/}).

\section{Conclusions}
In summary: we have identified two previously known young objects \priname\enspace and \secnameshort\enspace as having common proper motion. We find that the photometric distances of the pair agree and that they are unlikely to be an alignment of two unrelated young objects. Using the strength of the lithium 6708\AA\enspace feature we find an age range of 10--45\,Myr for \priname\enspace yielding a mass of 11.6--15$M_J$ for \secnameshort. We note that the system has a wider separation than any known star-planet system and that \secnameshort\enspace is similar in age, mass and temperature to the known exoplanet $\beta$~Pic~b.

\section*{Acknowledgments}
This research made use of the SIMBAD database, operated at CDS, Strasbourg, France. This publication makes use of data products from the Two Micron All Sky Survey, which is a joint project of the University of Massachusetts and the Infrared Processing and Analysis Center/California Institute of Technology, funded by the National Aeronautics and Space Administration and the National Science Foundation. This publication also makes use of data products from NEOWISE, which is a project of the Jet Propulsion Laboratory/California Institute of Technology, funded by the Planetary Science Division of the National Aeronautics and Space Administration.
The DENIS project has been partly funded by the SCIENCE and the HCM plans of
the European Commission under grants CT920791 and CT940627.
It is supported by INSU, MEN and CNRS in France, by the State of Baden-W\"urttemberg 
in Germany, by DGICYT in Spain, by CNR in Italy, by FFwFBWF in Austria, by FAPESP in Brazil,
by OTKA grants F-4239 and F-013990 in Hungary, and by the ESO C\&EE grant A-04-046. Jean Claude Renault from IAP was the Project manager.  Observations were  
carried out thanks to the contribution of numerous students and young 
scientists from all involved institutes, under the supervision of  P. Fouqu\'e,  
survey astronomer resident in Chile. Based on observations obtained at the Gemini Observatory (PID GS-2009B-C-2, acquired through the Gemini Science Archive), which is operated by the 
Association of Universities for Research in Astronomy, Inc., under a cooperative agreement 
with the NSF on behalf of the Gemini partnership: the National Science Foundation 
(United States), the National Research Council (Canada), CONICYT (Chile), the Australian 
Research Council (Australia), Minist\'{e}rio da Ci\^{e}ncia, Tecnologia e Inova\c{c}\~{a}o 
(Brazil) and Ministerio de Ciencia, Tecnolog\'{i}a e Innovaci\'{o}n Productiva (Argentina). The research of J.E.S. was supported by an appointment to the NASA Postdoctoral Program at NASA Ames Research Center, administered by Oak Ridge Associated Universities through a contract with NASA. We thank the anonymous referee for the prompt and helpful review that improved the quality and clarity of this manuscript. The authors would like to the thank the Brass Monkey, Heidelberg for the opportunity to pit their wits against each other every week.

\bibliography{ndeacon}

\begin{thebibliography}{52}
\expandafter\ifx\csname natexlab\endcsname\relax\def\natexlab#1{#1}\fi

\bibitem[{Allard {et~al}\mbox{.}(2011)Allard, Homeier, \& Freytag}]{Allard2010}
Allard F., Homeier D., Freytag B., 2011, in 16th Cambridge Work. Cool Stars.
  Stellar Syst. Sun, Johns-Krull C.~M., Browning M.~K., West A.~A., eds., ASP
  Conference Series Vol. 448, San Francisco, p.~91

\bibitem[{Allard {et~al}\mbox{.}(2012)Allard, Homeier, \& Freytag}]{Allard2012}
Allard F., Homeier D., Freytag B., 2012, Philos. Trans. R. Soc. A Math. Phys.
  Eng. Sci., 370, 2765

\bibitem[{Allers \& Liu(2013)}]{Allers2013}
Allers K.~N., Liu M.~C., 2013, Astrophys. J., 772, 79

\bibitem[{Baraffe {et~al}\mbox{.}(2003)Baraffe, Chabrier, Barman, Allard, \&
  Hauschildt}]{Baraffe2003}
Baraffe I., Chabrier G., Barman T.~S., Allard F., Hauschildt P.~H., 2003,
  Astron. Astrophys., 712, 701

\bibitem[{Baraffe {et~al}\mbox{.}(2015)Baraffe, Homeier, Allard, \&
  Chabrier}]{Baraffe2015}
Baraffe I., Homeier D., Allard F., Chabrier G., 2015, Astron. Astrophys., 577,
  A42

\bibitem[{Bell {et~al}\mbox{.}(2015)Bell, Mamajek, \& Naylor}]{Bell2015}
Bell C. P.~M., Mamajek E.~E., Naylor T., 2015, Mon. Not. R. Astron. Soc., 454,
  593

\bibitem[{Bonnefoy {et~al}\mbox{.}(2013)Bonnefoy, Boccaletti, Lagrange, Allard,
  Mordasini, Beust, Chauvin, Girard, Homeier, Apai, Lacour, \&
  Rouan}]{Bonnefoy2013}
Bonnefoy M. {et~al.}, 2013, Astron. Astrophys., 555, A107

\bibitem[{Bowler {et~al}\mbox{.}(2015)Bowler, Shkolnik, Liu, Schlieder, Mann,
  Dupuy, Hinkley, Crepp, Johnson, Howard, Flagg, Weinberger, Aller, Allers,
  Best, Kotson, Montet, Herczeg, Baranec, Riddle, Law, Nielsen, Wahhaj, Biller,
  \& Hayward}]{Bowler2015}
Bowler B.~P. {et~al.}, 2015, Astrophys. J., 806, 62

\bibitem[{Cruz {et~al}\mbox{.}(2009)Cruz, Kirkpatrick, \& Burgasser}]{Cruz2009}
Cruz K.~L., Kirkpatrick J.~D., Burgasser A.~J., 2009, Astron. J., 137, 3345

\bibitem[{da~Silva {et~al}\mbox{.}(2009)da~Silva, Torres, de~la Reza, Quast,
  Melo, \& Sterzik}]{DaSilva2009}
da~Silva L., Torres C. A.~O., de~la Reza R., Quast G.~R., Melo C. H.~F.,
  Sterzik M.~F., 2009, Astron. Astrophys., 508, 833

\bibitem[{Dopita {et~al}\mbox{.}(2007)Dopita, Hart, McGregor, Oates, Bloxham,
  \& Jones}]{Dopita2007}
Dopita M., Hart J., McGregor P., Oates P., Bloxham G., Jones D., 2007,
  Astrophys. Space Sci., 310, 255

\bibitem[{Elliott {et~al}\mbox{.}(2015)Elliott, Hu{\'{e}}lamo, Bouy, Bayo,
  Melo, Torres, Sterzik, Quast, Chauvin, \& Barrado}]{Elliott2015}
Elliott P. {et~al.}, 2015, Astron. Astrophys., 580, A88

\bibitem[{Epchtein {et~al}\mbox{.}(1994)Epchtein, {De Batz}, Copet,
  Fouqu{\'{e}}, Lacombe, {Le Bertre}, Mamon, Rouan, Tiph{\`{e}}ne, Burton,
  Deul, Habing, Borsenberger, Dennefeld, Omont, Renault, Volmerange,
  Kimeswenger, Appenzeller, Bender, Forveille, Garzon, Hron, Persi,
  Ferrari-Toniolo, \& Vauglin}]{Epchtein1994}
Epchtein N. {et~al.}, 1994, Astrophys. Space Sci., 217, 3

\bibitem[{Faherty {et~al}\mbox{.}(2013)Faherty, Rice, Cruz, Mamajek, \&
  N{\'{u}}{\~{n}}ez}]{Faherty2013}
Faherty J.~K., Rice E.~L., Cruz K.~L., Mamajek E.~E., N{\'{u}}{\~{n}}ez A.,
  2013, Astron. J., 145, 2

\bibitem[{Filippazzo {et~al}\mbox{.}(2015)Filippazzo, Rice, Faherty, Cruz, {Van
  Gordon}, \& Looper}]{Filippazzo2015}
Filippazzo J.~C., Rice E.~L., Faherty J., Cruz K.~L., {Van Gordon} M.~M.,
  Looper D.~L., 2015, Astrophys. J., 810, 158

\bibitem[{Gagn{\'{e}} {et~al}\mbox{.}(2015{\natexlab{a}})Gagn{\'{e}}, Faherty,
  Cruz, Lafreni{\'{e}}re, Doyon, Malo, Burgasser, Naud, Artigau, Bouchard,
  Gizis, \& Albert}]{Gagne2015a}
Gagn{\'{e}} J. {et~al.}, 2015{\natexlab{a}}, Astrophys. J. Suppl. Ser., 219, 33

\bibitem[{Gagn{\'{e}} {et~al}\mbox{.}(2014)Gagn{\'{e}}, Lafreni{\`{e}}re,
  Doyon, Malo, \& Artigau}]{Gagne2014}
Gagn{\'{e}} J., Lafreni{\`{e}}re D., Doyon R., Malo L., Artigau {\'{E}}., 2014,
  Astrophys. J., 783, 121

\bibitem[{Gagn{\'{e}} {et~al}\mbox{.}(2015{\natexlab{b}})Gagn{\'{e}},
  Lafreni{\`{e}}re, Doyon, Malo, \& Artigau}]{Gagne2015}
Gagn{\'{e}} J., Lafreni{\`{e}}re D., Doyon R., Malo L., Artigau {\'{E}}.,
  2015{\natexlab{b}}, Astrophys. J., 798, 73

\bibitem[{Gaidos {et~al}\mbox{.}(2014)Gaidos, Mann, Lepine, Buccino, James,
  Ansdell, Petrucci, Mauas, \& Hilton}]{Gaidos2014}
Gaidos E. {et~al.}, 2014, Mon. Not. R. Astron. Soc., 443, 2561

\bibitem[{Goldman {et~al}\mbox{.}(2010)Goldman, Marsat, Henning, Clemens, \&
  Greiner}]{Goldman2010}
Goldman B., Marsat S., Henning T., Clemens C., Greiner J., 2010, Mon. Not. R.
  Astron. Soc., 405, 1140

\bibitem[{Hinkle {et~al}\mbox{.}(2003)Hinkle, Blum, Joyce, Sharp, Ridgway,
  van~der Bliek, Rogers, Smith, \& Valenti}]{Hinkle2003}
Hinkle K.~H. {et~al.}, 2003, in SPIE 4834, Discov. Res. Prospect. from 6- to
  10-Meter-Class Telesc. II, Guhathakurta P., ed., pp. 353--363

\bibitem[{Kaufer {et~al}\mbox{.}(1999)Kaufer, Stahl, Tubbesing,
  N{\{}{\{}$\backslash$o{\}}{\}}rregaard, Avila, Francois, Pasquini, \&
  Pizzella}]{Kaufer1999}
Kaufer A., Stahl O., Tubbesing S., N{\{}{\{}$\backslash$o{\}}{\}}rregaard P.,
  Avila G., Francois P., Pasquini L., Pizzella A., 1999, The Messenger, 95, 8

\bibitem[{Kraus {et~al}\mbox{.}(2014)Kraus, Shkolnik, Allers, \&
  Liu}]{Kraus2014a}
Kraus A.~L., Shkolnik E.~L., Allers K.~N., Liu M.~C., 2014, Astron. J., 147,
  146

\bibitem[{Kuzuhara {et~al}\mbox{.}(2011)Kuzuhara, Tamura, Ishii, Kudo,
  Nishiyama, \& Kandori}]{Kuzuhara2011}
Kuzuhara M., Tamura M., Ishii M., Kudo T., Nishiyama S., Kandori R., 2011,
  Astron. J., 141, 119

\bibitem[{Lagrange {et~al}\mbox{.}(2010)Lagrange, Bonnefoy, Chauvin, Apai,
  Ehrenreich, Boccaletti, Gratadour, Rouan, Mouillet, Lacour, \&
  Kasper}]{Lagrange2010}
Lagrange A.-M. {et~al.}, 2010, Science, 329, 57

\bibitem[{L{\'{e}}pine \& Bongiorno(2007)}]{Lepine2007}
L{\'{e}}pine S., Bongiorno B., 2007, Astron. J., 133, 889

\bibitem[{L{\'{e}}pine {et~al}\mbox{.}(2013)L{\'{e}}pine, Hilton, Mann, Wilde,
  Rojas-Ayala, Cruz, \& Gaidos}]{Lepine2013}
L{\'{e}}pine S., Hilton E.~J., Mann A.~W., Wilde M., Rojas-Ayala B., Cruz
  K.~L., Gaidos E., 2013, Astron. J., 145, 102

\bibitem[{Luhman(2013)}]{Luhman2013}
Luhman K.~L., 2013, Astrophys. J., 767, L1

\bibitem[{Luhman {et~al}\mbox{.}(2011)Luhman, Burgasser, \&
  Bochanski}]{Luhman2011}
Luhman K.~L., Burgasser A.~J., Bochanski J.~J., 2011, Astrophys. J., 730, L9

\bibitem[{Luhman {et~al}\mbox{.}(2012)Luhman, Loutrel, McCurdy, Mace, Melso,
  Star, Young, Terrien, McLean, {Davy Kirkpatrick}, \& Rhode}]{Luhman2012}
Luhman K.~L. {et~al.}, 2012, Astrophys. J., 760, 152

\bibitem[{Mainzer {et~al}\mbox{.}(2011)Mainzer, Bauer, Grav, Masiero, Cutri,
  Dailey, Eisenhardt, McMillan, Wright, Walker, Jedicke, Spahr, Tholen, Alles,
  Beck, Brandenburg, Conrow, Evans, Fowler, Jarrett, Marsh, Masci, McCallon,
  Wheelock, Wittman, Wyatt, DeBaun, Elliott, Elsbury, Gautier, Gomillion,
  Leisawitz, Maleszewski, Micheli, \& Wilkins}]{Mainzer2011}
Mainzer A. {et~al.}, 2011, Astrophys. J., 731, 53

\bibitem[{Malo {et~al}\mbox{.}(2014)Malo, Artigau, Doyon, Lafreni{\`{e}}re,
  Albert, \& Gagn{\'{e}}}]{Malo2014}
Malo L., Artigau {\'{E}}., Doyon R., Lafreni{\`{e}}re D., Albert L.,
  Gagn{\'{e}} J., 2014, Astrophys. J., 788, 81

\bibitem[{Malo {et~al}\mbox{.}(2013)Malo, Doyon, Lafreni{\`{e}}re, Artigau,
  Gagn{\'{e}}, Baron, \& Riedel}]{Malo2013}
Malo L., Doyon R., Lafreni{\`{e}}re D., Artigau {\'{E}}., Gagn{\'{e}} J., Baron
  F., Riedel A., 2013, Astrophys. J., 762, 88

\bibitem[{Manjavacas {et~al}\mbox{.}(2014)Manjavacas, Bonnefoy, Schlieder,
  Allard, Rojo, Goldman, Chauvin, Homeier, Lodieu, \& Henning}]{Manjavacas2014}
Manjavacas E. {et~al.}, 2014, Astron. Astrophys., 564, A55

\bibitem[{Martin {et~al}\mbox{.}(2005)Martin, Fanson, Schiminovich, Morrissey,
  Friedman, Barlow, Conrow, Grange, Jelinsky, Milliard, Siegmund, Bianchi,
  Byun, Donas, Forster, Heckman, Lee, Madore, Malina, Neff, Rich, Small,
  Surber, Szalay, Welsh, \& Wyder}]{Martin2005}
Martin D.~C. {et~al.}, 2005, Astrophys. J., 619, L1

\bibitem[{Mentuch {et~al}\mbox{.}(2008)Mentuch, Brandeker, van Kerkwijk,
  Jayawardhana, \& Hauschildt}]{Mentuch2008}
Mentuch E., Brandeker A., van Kerkwijk M.~H., Jayawardhana R., Hauschildt
  P.~H., 2008, Astrophys. J., 689, 1127

\bibitem[{Morzinski {et~al}\mbox{.}(2015)Morzinski, Males, Skemer, Close, Hinz,
  Rodigas, Puglisi, Esposito, Riccardi, Pinna, Xompero, Briguglio, Bailey,
  Follette, Kopon, Weinberger, \& Wu}]{Morzinski2015}
Morzinski K.~M. {et~al.}, 2015

\bibitem[{Murphy \& Lawson(2014)}]{Murphy2014}
Murphy S.~J., Lawson W.~A., 2014, Mon. Not. R. Astron. Soc., 447, 1267

\bibitem[{Naud {et~al}\mbox{.}(2014)Naud, Artigau, Malo, Albert, Doyon,
  Lafreni{\`{e}}re, Gagn{\'{e}}, Saumon, Morley, Allard, Homeier, Beichman,
  Gelino, \& Boucher}]{Naud2014}
Naud M.-E. {et~al.}, 2014, Astrophys. J., 787, 5

\bibitem[{Pecaut \& Mamajek(2013)}]{Pecaut2013}
Pecaut M.~J., Mamajek E.~E., 2013, Astrophys. J. Suppl. Ser., 208, 9

\bibitem[{Prato {et~al}\mbox{.}(2002)Prato, Simon, Mazeh, McLean, Norman, \&
  Zucker}]{Prato2002}
Prato L., Simon M., Mazeh T., McLean I.~S., Norman D., Zucker S., 2002,
  Astrophys. J., 569, 863

\bibitem[{Reid {et~al}\mbox{.}(2008)Reid, Cruz, Kirkpatrick, Allen, Mungall,
  Liebert, Lowrance, \& Sweet}]{Reid2008}
Reid I., Cruz K.~L., Kirkpatrick J.~D., Allen P.~R., Mungall F., Liebert J.,
  Lowrance P., Sweet A., 2008, Astron. J., 136, 1290

\bibitem[{Saumon \& Marley(2008)}]{Saumon2008}
Saumon D., Marley M.~S., 2008, Astrophys. J., 689, 1327

\bibitem[{Schlieder {et~al}\mbox{.}(2012)Schlieder, L{\'{e}}pine, \&
  Simon}]{Schlieder2012}
Schlieder J.~E., L{\'{e}}pine S., Simon M., 2012, Astron. J., 143, 80

\bibitem[{Shkolnik {et~al}\mbox{.}(2009)Shkolnik, Liu, \& Reid}]{Shkolnik2009}
Shkolnik E., Liu M.~C., Reid I.~N., 2009, Astrophys. J., 699, 649

\bibitem[{Skrutskie {et~al}\mbox{.}(2006)Skrutskie, Cutri, Stiening, Weinberg,
  Schneider, Carpenter, Beichman, Capps, Chester, Elias, Huchra, Liebert,
  Lonsdale, Monet, Price, Seitzer, Jarrett, Kirkpatrick, Gizis, Howard, Evans,
  Fowler, Fullmer, Hurt, Light, Kopan, Marsh, McCallon, Tam, {Van Dyk}, \&
  Wheelock}]{Skrutskie2006}
Skrutskie M.~F. {et~al.}, 2006, Astron. J., 131, 1163

\bibitem[{Thomas {et~al}\mbox{.}(1998)Thomas, Beuermann, Reinsch, Schwope,
  Truemper, \& Voges}]{Thomas1998}
Thomas H.-C., Beuermann K., Reinsch K., Schwope A.~D., Truemper J., Voges W.,
  1998, Astron. Astrophys., 335, 467

\bibitem[{Torres {et~al}\mbox{.}(2006)Torres, Quast, da~Silva, de~la Reza,
  Melo, \& Sterzik}]{Torres2006}
Torres C. A.~O., Quast G.~R., da~Silva L., de~la Reza R., Melo C. H.~F.,
  Sterzik M., 2006, Astron. Astrophys., 460, 695

\bibitem[{Torres {et~al}\mbox{.}(2008)Torres, Quast, Melo, \&
  Sterzik}]{Torres2008}
Torres C. A.~O., Quast G.~R., Melo C. H.~F., Sterzik M.~F., 2008, in Handb.
  Star Form. Reg. Vol. II South. Sky, Reipurth B., ed., ASP Monograph
  Publications, Vol. 5., p. 757

\bibitem[{Wahhaj {et~al}\mbox{.}(2011)Wahhaj, Liu, Biller, Clarke, Nielsen,
  Close, Hayward, Mamajek, Cushing, Dupuy, Tecza, Thatte, Chun, Ftaclas,
  Hartung, Reid, Shkolnik, Alencar, Artymowicz, Boss, {de Gouveia Dal Pino},
  Gregorio-Hetem, Ida, Kuchner, Lin, \& Toomey}]{Wahhaj2011}
Wahhaj Z. {et~al.}, 2011, Astrophys. J., 729, 139

\bibitem[{Wright {et~al}\mbox{.}(2010)Wright, Eisenhardt, Mainzer, Ressler,
  Cutri, Jarrett, Kirkpatrick, Padgett, McMillan, Skrutskie, Stanford, Cohen,
  Walker, Mather, Leisawitz, Gautier, McLean, Benford, Lonsdale, Blain, Mendez,
  Irace, Duval, Liu, Royer, Heinrichsen, Howard, Shannon, Kendall, Walsh,
  Larsen, Cardon, Schick, Schwalm, Abid, Fabinsky, Naes, \& Tsai}]{Wright2010}
Wright E.~L. {et~al.}, 2010, Astron. J., 140, 1868

\bibitem[{Zacharias {et~al}\mbox{.}(2013)Zacharias, Finch, Girard, Henden,
  Bartlett, Monet, \& Zacharias}]{Zacharias2013}
Zacharias N., Finch C.~T., Girard T.~M., Henden A., Bartlett J.~L., Monet
  D.~G., Zacharias M.~I., 2013, Astron. J., 145, 44

\end{thebibliography}

\bibliographystyle{mn2e}
\label{lastpage}

\end{document}